\documentclass
[superscriptaddress,secnumarabic,amssymb,amsmath,nobibnotes,aps,prd,showkeys,showpacs,nofootinbib,onecolumn,12pt]{revtex4}%
\usepackage{graphicx}
\usepackage{epsf}
\usepackage{bm}
\usepackage{amsmath}
\usepackage{amsfonts}
\usepackage{amssymb}%
\providecommand{\U}[1]{\protect\rule{.1in}{.1in}}

\newcommand{\be}{\begin{equation}}
\newcommand{\ee}{\end{equation}}

\newcommand{\mincir}{\raise
-3.truept\hbox{\rlap{\hbox{$\sim$}}\raise4.truept\hbox{$<$}\ }}
\newcommand{\magcir}{\raise
-3.truept\hbox{\rlap{\hbox{$\sim$}}\raise4.truept\hbox{$>$}\ }}

\begin{document}
\title{$O\left(  d,d\right)  $ symmetry in teleparallel dark energy}
\author{Andronikos Paliathanasis}
\email{anpaliat@phys.uoa.gr}
\affiliation{Institute of Systems Science, Durban University of Technology, Durban 4000,
South Africa }
\affiliation{Instituto de Ciencias F\'{\i}sicas y Matem\'{a}ticas, Universidad Austral de
Chile, Valdivia 5090000, Chile}

\begin{abstract}
An important characteristic of the Dilaton cosmological model is the
Gasperini-Veneziano duality transformation which follows from the existence of
the $O\left(  d,d\right)  $ symmetry. In this study, we consider the
equivalent Dilaton theory in teleparallel dark energy with the $O\left(
d,d\right)  $ symmetry, while the equivalent teleparallel-duality
transformation is presented. The classical solution of the field equations is
derived. Finally the Wheeler-DeWitt equation of quantum cosmology is discussed.

\end{abstract}
\date{\today}
\maketitle

\section{Introduction}

\label{sec1}

A main property of the conformal field theory is the duality symmetry. In
\cite{gas1}, Veneziano introduced the duality symmetry in string theory in
order to solve problem in the early universe. This approach opened a new
subject of study known as string cosmology. String cosmology is based on the
existence of a scalar field scalar field $\phi\left(  x^{k}\right)  ,~$known
as the dilaton field,~coupled to gravity.\ The gravitational Action integral
is defined as \cite{s1,s2,s3}%
\begin{equation}
S_{dilaton}=\int d^{D}x\sqrt{-g}e^{-2\phi}\left(  R-4g^{\mu\nu}\phi_{;\mu}%
\phi_{;\nu}+\Lambda\right)  , \label{d.01}%
\end{equation}
where we have assumed the antisymmetric tensor strength of the sigma model,
constructed by the three-form axion fields to be zero. $\Lambda$ is the
cosmological constant term, and $R$ is the Ricciscalar of the background
space~with metric tensor $g_{\mu\nu}$. The Action Integral (\ref{d.01}) has
many similarities with the Brans-Dicke theory \cite{bd}. Indeed, the
Brans-Dicke theory for a fixed Brans-Dicke parameter is recovered after the
change of variables $\phi\left(  x^{k}\right)  =-\frac{1}{2}\ln\psi\left(
x^{k}\right)  ~$\cite{an1}.

The main properties of string cosmology are summarized in \cite{gas2}.
Specifically, inflation in string cosmology follows naturally without impose
any fine-tune potential, the electromagnetic perturbations can describe the
galactic magnetic fields, while matter perturbations remain small to support
the homogeneity of the universe, for more details we refer the reader to
\cite{sr1}.

In the case of a $D$-dimensional spatially flat and homogeneous background
space, the Lagrangian function for the dilaton field depends only the scale
factor $a\left(  t\right)  ~$and on the scalar field~$\phi\left(  t\right)  $.
The Lagrangian function it admits the scale-factor duality property, that is,
the Action Integral (\ref{d.01}) is invariant under the transformation
\cite{gas1}%
\begin{equation}
a\left(  t\right)  \rightarrow a^{-1}\left(  t\right)  ~,~\phi\left(
t\right)  \rightarrow\phi\left(  t\right)  -\left(  D-1\right)  \ln
a.\label{d.02}%
\end{equation}

The scale-factor duality transformation has been generalized and for the case
of anisotropic and inhomogeneous spacetimes in \cite{s1}. Nowadays it is known
as Gasperini-Veneziano duality property. The duality symmetry is a discrete
transformation and an isometry should exist in the background space
\cite{s2,s3}. The fundamental origin for the Gasperini-Veneziano
transformation\ is the $O\left(  d,d\right)  $ symmetry \cite{s1}.
Furthermore, for the Hubble function in a spatially flat
Friedmann--Lema\^{\i}tre--Robertson--Walker (FLRW) space, the duality
transformation reads $H\rightarrow-H$. Therefore, under the second discrete
transformation $t\rightarrow-t$, it follows, $\dot{H}\left(  t\right)
\rightarrow\dot{H}\left(  -t\right)  $ which leads to the string-driven
pre-big-bang cosmology \cite{gas3}. Furthermore, it was found that the dilaton
field (\ref{d.01}) admits a large number of symmetries which were used to
solve completely the Wheeler-DeWitt equation of quantum cosmology \cite{keh}.
In the case of cosmological studies, the scale-factor duality transformation
has been attributed to a local transformation which leave invariant the Action
Integral (\ref{d.01}), that is, a variational symmetry \cite{an1}. Moreover,
the viability of the discrete transformation under conformal transformations
investigated in \cite{ton2}. In addition, a new mathematical construction
approach for the determination of discrete transformations established in
\cite{ton2}

In this piece of work, we are interested on the existence of discrete
transformations, similar with the scale-factor duality transformation
(\ref{d.02}), in the case of teleparallel dark energy theory \cite{tde}. The
aforementioned theory belongs to the so-called alternative/modified theories
of gravity \cite{fg2,fg3,fg4,fg11,fg15,fg20,ftb02,md1,md2,md3} which have been
introduced the last years by cosmologists in order to explain the cosmological
observations \cite{sp1,fg5}. Teleparallelism has drawn attention the last
years because it provides a systematic geometric description for the
explanation of the cosmological observations. In the teleparallel equivalence
of General Relativity \cite{ein28}, instead of the torsion-less Levi-Civita
connection the curvatureless Weitzenb\"{o}ck connection in considered, where
the corresponding dynamical fields are the four linearly independent
vierbeins. Hence, the gravitational field are defined by the Weitzenb\"{o}ck
tensor and its scalar $T~$\cite{Hayashi79,Tsamp,ft1}. The teleparallel dark
energy is the analogue of the scalar tensor theories, where a scalar field is
introduced in teleparallel Action Integral while the scalar field interacts
with the scalar $T$ for the Weitzenb\"{o}ck tensor. The theory is also known
as scalar-torsion theory \cite{sss1,ss2,ss3,ss4}. Under a conformal
transformation \cite{ct1} the theory can be related with the so-called
$f\left(  T,B\right)  $ theory \cite{ct2}, which is a fourth-order theory of
gravity, as an analogue of the equivalence of O'\ Hanlon theory with $f\left(
R\right)  $ gravity \cite{ct3}. The evolution of the cosmological dynamics in
teleparallel dark energy was studied before in \cite{te1,te2,te3}. Analysis of
the cosmological observations with the teleparallel dark energy are presented
in \cite{te4,te5} while some other studies are given in \cite{te6,te7}. From
the latter studies it is clear that the theory can play an important role for
the description of the late-time and the early acceleration phases of the
universe. Moreover, from the analysis of the evolution for the matter
perturbations in teleparallel dark energy \ref{te5} it was found that
teleparallel dark energy theory is favoured with respect to the quintessence
theory. \ Modified Teleparallel theories of gravity are Lorentz violated
theories. Nowadays Lorentz violation has not been observed; however, Lorentz
violation is a prediction for various models of quantum gravity, for a review
see \cite{dav1}. The motivation of this work is to define a teleparallel dark
energy model which admits a discrete transformation and open the way for the
teleparallel string cosmology. The plan of the paper is as follows.

In Section \ref{sec2}, we present the field equations for the teleparallel
dark energy. In\ Section \ref{sec3} we define the teleparallel dilaton model,
which is invariant under a discrete transformation similar to the scale-factor
duality transformation of the dilaton cosmological model. This new discrete
transformation it has its origin on the presence of the $O\left(  d,d\right)
$ symmetry. Moreover, the field equations are found to be superintegrable and
the analytic solution is expressed in terms of exponential functions. Finally,
in Section \ref{con00} we summarize our results, while we solve the
Wheeler-DeWitt equation of quantum cosmology for the teleparallel dilaton model.

\section{Teleparallel dark energy}

\label{sec2}

In teleparallelism the dynamical variables are the vierbein fields.\ They are
defined by the requirement $g(e_{i},e_{j})=e_{i}.e_{j}=\eta_{ij}$~where
$\eta_{ij}=\mathrm{diag}(1,-1,-1,-1)$ is the Lorentz metric in canonical form.

The metric tensor $g_{\mu\nu}(x^{\kappa})$ in terms of coordinates is defined
as%
\begin{equation}
g_{\mu\nu}=\eta_{ij}h_{~\mu}^{i}h_{~\nu}^{j},\label{d.04}%
\end{equation}
where $e^{i}(x^{\kappa})=h_{\mu}^{i}(x^{\kappa})dx^{i}$ is the dual basis, in
which $e^{i}\left(  e_{j}\right)  =\delta_{j}^{i}$. 

In contrary to General Relativity the, curvatureless teleparallel torsion
tensor is the fundamental geometric object in teleparallelism, is defined by
the  antisymmetric part of the affine connection coefficients as follows
\begin{equation}
T_{\mu\nu}^{~~\beta}=\hat{\Gamma}_{\nu\mu}^{\beta}-\hat{\Gamma}_{\mu\nu
}^{\beta}=h_{~~i}^{\beta}(\partial_{\mu}h_{~\nu}^{i}-\partial_{\nu}h_{~\mu
}^{i}).\label{d.05}%
\end{equation}
The gravitational Lagrangian for the teleparallel equivalent of General
Relativity is defined by the scalar \ $T=S_{~\beta}^{\mu\nu}T_{~\mu\nu}%
^{\beta}~$where~$S_{~\beta}^{\mu\nu}=\frac{1}{2}(K_{~\beta}^{\mu\nu}%
+\delta_{~\beta}^{\mu}T_{~\theta}^{\theta\nu}-\delta_{\beta}^{\nu}T_{~\theta
}^{\theta\mu})~$and $K_{~\beta}^{\mu\nu}$ is the tensor~$K_{~\beta}^{\mu\nu
}=-\frac{1}{2}(T_{~\beta}^{\mu\nu}-T_{~\beta}^{\nu\mu}-T_{~\beta}^{\mu\nu})$.
The latter tensor, equals the difference of the Levi Civita connection in the
holonomic and the unholonomic frame.

As stated by the cosmological principle the universe in large scales is
homogeneous and isotropic described by the spatially flat FLRW metric%
\begin{equation}
ds^{2}=N^{2}\left(  t\right)  dt^{2}-a^{2}(t)(dx^{2}+dy^{2}+dz^{2}%
).\label{d.06}%
\end{equation}
Therefore, in order to recover such cosmological scenario we assume the
diagonal frame for the vierbein fields $~h_{~\mu}^{i}(t)=\mathrm{diag}%
(1,a(t),a(t),a(t))$, where we calculate the scalar
\begin{equation}
T=6H^{2},\label{d.07}%
\end{equation}
in which $H=\frac{1}{N}\frac{\dot{a}}{a}$ is the Hubble function.

The gravitational Action Integral in teleparallel dark energy theory is
defined to be
\begin{equation}
S=\frac{1}{16\pi G}\int d^{4}xe\left[  F\left(  \phi\right)  \left(
T+\frac{\omega}{2}\phi_{;\mu}\phi^{\mu}+V\left(  \phi\right)  \right)
\right]  .\label{d.08}%
\end{equation}
where~$e=\sqrt{-g}$,~ $F\left(  \phi\right)  $ is the coupling function,
$\omega$ is a constant nonzero parameter, analogue of the Brans-Dicke
parameter and $V\left(  \phi\right)  $ is the scalar field potential. We
remark that we can always define new scalar field under the point
transformation $d\psi=\sqrt{\omega F\left(  \phi\right)  }d\phi$, such that
the Action Integral (\ref{d.08}) to be written as follows%
\begin{equation}
S=\frac{1}{16\pi G}\int d^{4}xe\left[  \hat{F}\left(  \psi\right)  T+\frac
{1}{2}\psi_{;\mu}\psi^{\mu}+\hat{V}\left(  \psi\right)  \right]  .\label{d.09}%
\end{equation}

The field equations in this cosmological model admit a minisuperspace
description. Indeed, by replace scar $T$ from (\ref{d.07}) in (\ref{d.08}) and
assume that the scalar field $\phi$ inherits the symmetries of the background
space, i.e. $\phi=\phi\left(  t\right)  $, the point-like Lagrangian for the
field equations is
\begin{equation}
L\left(  a,\dot{a},\phi,\dot{\phi}\right)  =F\left(  \phi\right)  \left(
\frac{1}{N}\left(  6a\dot{a}^{2}-\frac{\omega}{2}a^{3}\dot{\phi}^{2}\right)
+Na^{3}V\left(  \phi\right)  \right)  .\label{d.10}%
\end{equation}

Hence, the gravitational field equations are derived by the variation of the
Lagrangian function (\ref{d.10}). Indeed, the field equations are
\begin{equation}
F\left(  \phi\right)  \left(  6H^{2}-\frac{\omega}{2N^{2}}\dot{\phi}%
^{2}-V\left(  \phi\right)  \right)  =0~,\label{d.11}%
\end{equation}%
\begin{equation}
\left(  \frac{2}{N}\dot{H}+3H^{2}\right)  +\frac{1}{2}\left(  \frac{\omega
}{2N^{2}}\dot{\phi}^{2}-V\left(  \phi\right)  \right)  +2\left(  \ln F\left(
\phi\right)  \right)  _{,\phi}H\frac{\dot{\phi}}{N}=0~,\label{d.12}%
\end{equation}%
\begin{equation}
\omega\left(  \frac{1}{N^{2}}\ddot{\phi}+\frac{3}{N}H\dot{\phi}-\frac{\dot{N}%
}{N^{3}}\dot{\phi}\right)  +\left(  \frac{\omega}{2N^{2}}\dot{\phi}%
^{2}+V\left(  \phi\right)  \right)  +V_{,\phi}\left(  \phi\right)
=0~.\label{d.13}%
\end{equation}

We continue our analysis by assuming specific functional forms for the
coupling function $F\left(  \phi\right)  $ and the potential $V\left(
\phi\right)  $ in which the field equations remain invariant under discrete
transformations as that of the scale-factor duality transformation for the
dilaton field. Without loss of generality in the following we assume the lapse
function to be constant, i.e. $N\left(  t\right)  =1$. In this case, equation
(\ref{d.11}) can be seen as a the constraint equation of the Hamiltonian for
the second-order differential equations (\ref{d.12}) and (\ref{d.13}).

\section{$O\left(  d,d\right)  $ symmetry in teleparallel dark energy}

\label{sec3}

For the unknown functions of the point-like Lagrangian (\ref{d.10}), that is,
the coupling function and the potential we consider that they are $F\left(
\phi\right)  =e^{-2\phi}$ and $V\left(  \phi\right)  =\Lambda$. Hence, the
point-like Lagrangian (\ref{d.10}) reads
\begin{equation}
L\left(  a,\dot{a},\phi,\dot{\phi}\right)  =e^{-2\phi}\left(  6a\dot{a}%
^{2}-\frac{\omega}{2}a^{3}\dot{\phi}^{2}+a^{3}\Lambda\right)  .\label{d.14}%
\end{equation}
This cosmological model we shall call it the teleparallel dilaton model, or
dilaton-tensor model. As we shall see in the following, for this specific
selection of the free functions, the point-like Lagrangian (\ref{d.14}) is
invariant under the $O\left(  d,d\right)  $ symmetry. We continue with the
construction of the discrete transformation and the derivation of the
$O\left(  d,d\right)  $ symmetry for the field equations. 

We observe that under the scale factor duality transformation (\ref{d.02}),
for $D=4$, Lagrangian function (\ref{d.14}) does not remain invariant. Thus,
we should investigate for other forms for the discrete transformation.

However, we observe that under the change of variables
\begin{equation}
a\rightarrow\bar{a}^{p_{1}}e^{p_{2}\bar{\phi}}~,~\phi\rightarrow p_{4}%
\bar{\phi}+p_{3}\ln a \label{d.15}%
\end{equation}
with%
\begin{equation}
p_{1}=\frac{1+\kappa^{2}}{1-\kappa^{2}}~,~p_{2}=-\frac{4}{3\left(
1-\kappa^{2}\right)  }~,~p_{3}=\frac{3\kappa^{2}}{1-\kappa^{2}}~,~p_{4}%
=-\frac{1+\kappa^{2}}{1-\kappa^{2}}~,~\omega=\frac{4}{3\kappa^{2}}.
\label{d.16}%
\end{equation}
the Lagrangian function (\ref{d.14}) becomes%
\begin{equation}
L\left(  a,\dot{a},\phi,\dot{\phi}\right)  =e^{-2\bar{\phi}}\left(  6\bar
{a}\left(  \frac{d\bar{a}}{dt}\right)  ^{2}-\frac{\omega}{2}\bar{a}^{3}\left(
\frac{d\bar{\phi}}{dt}\right)  ^{2}+a^{3}\Lambda\right)  . \label{d.17}%
\end{equation}
Thus, the discrete transformation (\ref{d.15}) with (\ref{d.16}) is a symmetry
for the teleparallel dilaton model. In the case of large values of $\kappa$,
the discrete transformation (\ref{d.15}) becomes~$a\rightarrow\bar{a}%
^{-1}~,~\phi\rightarrow\bar{\phi}-3\ln\bar{a}~$which is the
Gasperini-Veneziano scale factor duality. Furthermore, for $\omega=\frac{8}%
{3}$, that is, $\kappa^{2}=1$, the discrete transformation does not exist.

In order to understand the origin of this discrete transformation, consider
the point transformation%
\begin{align}
u\left(  a,\phi;\kappa\right)   &  =\frac{8}{3\left(  1+\kappa\right)
}a^{\frac{3}{2}\left(  1+\kappa\right)  }\exp\left(  -\frac{1+\kappa}{\kappa
}\phi\right)  ~,~\label{d.18}\\
v\left(  a,\phi;\kappa\right)   &  =-\frac{1}{1-\kappa}a^{\frac{3}{2}\left(
1-\kappa\right)  }\exp\left(  \frac{1-\kappa}{\kappa}\phi\right)  .
\label{d.19}%
\end{align}
Therefore, in the new variables the point-like Lagrangian (\ref{d.14}) becomes%
\begin{equation}
L\left(  u,v,\dot{u},\dot{v}\right)  =-\left(  \dot{u}\dot{v}+\frac{3}%
{8}\left(  1-\kappa^{2}\right)  \Lambda uv\right)  . \label{d.20}%
\end{equation}
Hence, the discrete transformation (\ref{d.15}) in the new variables becomes
$\left\{  x\rightarrow\bar{y}~,~y\rightarrow\bar{x}\right\}  $, which is the
rotational symmetry in the two dimensional plane, i.e. the origin of
(\ref{d.15}) is the $O\left(  d,d\right)  $ symmetry.

Moreover, in the new variables, from (\ref{d.20}) the field equations reads%
\begin{align}
\dot{u}\dot{v}-\frac{3}{8}\left(  1-\kappa^{2}\right)  \Lambda uv  &
=0,\label{d.21}\\
\ddot{u}-\frac{3}{8}\left(  1-\kappa^{2}\right)  \Lambda u  &  =0,\label{d.22}%
\\
\ddot{v}-\frac{3}{8}\left(  1-\kappa^{2}\right)  \Lambda v  &  =0.
\label{d.23}%
\end{align}
The latter system is the two-dimensional oscillator, a well-known
superintegrable system. Hence, similarly with the dilaton field \cite{keh},
the existence of $O\left(  d,d\right)  $ symmetry in the teleparallel dark
energy theory leads to a superintegrable system.

The closed-form solution is
\begin{align}
u\left(  t\right)   &  =c_{1}e^{\sqrt{\bar{\Lambda}}t}+c_{2}e^{-\sqrt
{\bar{\Lambda}}t}~,\\
v\left(  t\right)   &  =c_{3}e^{\sqrt{\bar{\Lambda}}t}+c_{4}e^{-\sqrt
{\bar{\Lambda}}t}~,
\end{align}
with constraint $c_{1}c_{4}+c_{2}c_{3}=0$ and $\bar{\Lambda}=\frac{3}%
{8}\left(  1-\kappa^{2}\right)  \Lambda$. For large values of $t$ and for
$\bar{\Lambda}>0$, the solution becomes asymptotically $u\left(  t\right)
\simeq e^{\sqrt{\bar{\Lambda}}t}$,~\thinspace$v\left(  t\right)  \simeq
e^{\sqrt{\bar{\Lambda}}t}$, where it is clear that the final solution the
scale factor is $a\left(  t\right)  \simeq e^{\Omega\left(  \Lambda
,\kappa\right)  t}$, where $\Omega\left(  \Lambda,\kappa\right)  $ is a
constant. We conclude that de Sitter inflation is natural in teleparallel
dilaton theory.

Finally, for the Hubble function~$H\left(  t\right)  $ we find that under the
discrete transformation (\ref{d.15}) is transformed as $H\left(  t\right)
\rightarrow p_{1}\bar{H}\left(  t\right)  +p_{2}\dot{\phi}\left(  t\right)
$,~$\bar{H}\left(  t\right)  =\frac{d}{dt}\left(  \ln\bar{a}\right)  ,$ where
it is clear that for small values of $\omega$, $H\left(  t\right)
\rightarrow-\bar{H}\left(  t\right)  .$

\section{Conclusions}

\label{con00}

In this study, we generalized the dilaton\ cosmological model in the context
of teleparallel dark energy theory. For our new model, we developed that the
main properties of the dilaton field, i.e. the de Sitter inflation, and the
superintegrable property for the field equations, hold and for the
teleparallel dilaton model. The two theories share a common property, they
admit an isometry which is the $O\left(  d,d\right)  $ symmetry.

For the teleparallel dilaton model, we determined a discrete transformation
for the dynamical variables of the field equations, i.e. the scale factor
$a\left(  t\right)  $ and the scalar field $\phi\left(  t\right)  $, in which
the field equations remain invariant. This transformation is more general than
the Gasperini-Veneziano scale-factor duality transformation. While, the
Gasperini-Veneziano transformation is recovered in the case of the
teleparallel dilaton field when a free parameter for the model is very small.

As far as the Hubble function is concerned, the discrete symmetry in terms of
the Hubble function reads $H\left(  t\right)  \rightarrow p_{1}\left(
\kappa\right)  \bar{H}\left(  t\right)  +p_{2}\left(  \kappa\right)  \dot
{\phi}\left(  t\right)  $. Thus, for specific values of $\kappa$, the sign of
$H\left(  t\right)  $ can be changed such that under the second change of
variables $t\rightarrow-t$ ,\ we are able to study the pre-big bang epoch for
the universe in a similar way as in string cosmology. However, because of the
presence of the nonzero parameter $p_{2}\left(  \kappa\right)  $, the
behaviour in the pre-big bang epoch in the teleparallel model is different
from that of the dilaton field.

Finally, because of the existence of the minisuperspace Lagrangian
(\ref{d.14}) we are able to write the Wheeler-DeWitt equation of quantum
cosmology \cite{wd1}, similarly with the analysis presented in \cite{keh}. The
Hamiltonian constraint for the teleparallel dilaton model is written
\begin{equation}
\mathcal{H}\equiv e^{2\phi}\left(  \frac{p_{a}^{2}}{6a}-\frac{p_{\phi}^{2}%
}{2\omega a^{3}}-a^{3}\Lambda\right)  =0
\end{equation}
where the Wheeler-DeWitt equation reads~$\mathcal{W}\equiv\mathcal{H}\Psi,$
that is,%
\begin{equation}
\mathcal{W}\equiv e^{2\phi}\left(  \frac{1}{6a}\frac{\partial^{2}}{\partial
a^{2}}-\frac{1}{2\omega a^{3}}\frac{\partial^{2}}{\partial\phi^{2}}%
-a^{3}\Lambda\right)  \Psi\left(  a,\phi\right)  =0
\end{equation}

In the variables $\left\{  u,v\right\}  $ defined by expressions (\ref{d.18}),
(\ref{d.19}), the Wheeler-DeWitt equation is written in the simplest form%
\begin{equation}
\mathcal{W}\equiv\left(  \frac{\partial^{2}}{\partial u\partial v}%
-\bar{\Lambda}uv\right)  \Psi\left(  u,v\right)  =0.\label{d.24}%
\end{equation}
Equation (\ref{d.24}) admits the quantum operator the $\left(  \frac{\partial
}{\partial u^{2}}-\frac{\partial}{\partial v^{2}}-\bar{\Lambda}\left(
u^{2}-v^{2}\right)  \right)  \Psi=Q_{0}\Psi$, which is nothing else than the
Schr\"{o}dinger equation for the two-dimensional (hyperbolic) oscillator. Thus
we stop our discussion here.

We showed that the teleparallel dilaton model has important characteristics
similar to the classical dilaton model. That makes the model of special
interests for future studies. Furthermore, the generalization of the new
discrete transformation in the case of a anisotropic and inhomogeneous
background space should be investigated.

\end{document}